\pdfoutput=1
\documentclass[conference]{IEEEtran}
\usepackage{cite}
\usepackage{amsmath,amssymb,amsfonts}
\usepackage{algorithmic}
\usepackage{graphicx}
\usepackage{textcomp}
\usepackage{xcolor}
\usepackage{gensymb}
\usepackage{subfig}
\def\BibTeX{{\rm B\kern-.05em{\sc i\kern-.025em b}\kern-.08em
    T\kern-.1667em\lower.7ex\hbox{E}\kern-.125emX}}
\begin{document}

\title{Subband Beamforming in Coherent Hybrid Massive MIMO Using Eigenbeams}
\author{\IEEEauthorblockN{Chris Ng, Mihai Banu}
\IEEEauthorblockA{Blue Danube Systems, Warren, New Jersey, USA\\
chris.ng@bluedanube.com, mihaibanu@bluedanube.com}
}

\maketitle

\begin{abstract}
Hybrid Massive MIMO reduces implementation complexity but only supports beamforming coefficients that are common across all subbands.
However, in macro cellular where the channel has limited degrees of freedom, the long-term component of the channel can be decomposed into a set of subband-independent beamforming basis functions referred to as \emph{eigenbeams}.
A Coherent Hybrid Massive MIMO system can form arbitrary linear combinations of the eigenbeams at every subband to mimic Digital Massive MIMO beamforming as observed across all locations in the cell.
\end{abstract}

\section{Introduction}
In wireless cellular networks, the use of Massive MIMO \cite{marzetta:2010, marzetta:2016} is viewed as one of the key technologies to provide the next-generation of performance improvement.
MIMO refers to multiple-input multiple-output, or the use of multiple transmit and receive antennas in wireless communications.
In a Massive MIMO system, the base station is equipped with a large number of antenna elements, typically in an antenna array form factor.
A Digital Massive MIMO system has an independent radio frequency (RF) chain associated with each antenna element, and it is able to set different beamforming coefficients for different subbands.
Recently, Hybrid Massive MIMO architecture \cite{molisch:2017, vaghefi:2018} has gained much attention, since it reduces implementation complexity by having reduced number of RF chains and using analog circuitry to set the beamforming coefficients.
Typically, the analog attenuators and phase shifters cannot be set in a subband-specific manner.
As such, it has been perceived that Hybrid Massive MIMO systems cannot perform subband beamforming.

In this paper, we show that in macro cellular, the degrees of freedom of the channel is limited.
Under this environment, a Coherent Hybrid Massive MIMO system can be used as a \emph{radiation pattern synthesizer} across all subbands, by forming linear combinations at each subband of a set of analog beam coefficients we refer to as \emph{eigenbeams}.
The eigenbeams are effectively a set of basis functions induced by the long-term channel of the environment to allow the Coherent Hybrid Massive MIMO system mimic its digital counterpart at each subband as observed across the whole cell.

\section{System Model}

\subsection{Channel Model}
\label{sec:channel_model}

\begin{figure}[htbp]
\centerline{\includegraphics[width=\linewidth,trim={0 10cm 0 0cm},clip]{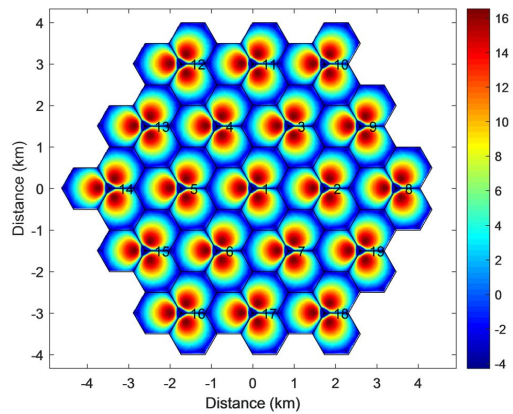}}
\caption{Typical macro cellular geometry and SINR (dB) distribution.}
\label{fig:cell_sinr}
\end{figure}

\begin{figure}[htbp]
\centerline{\includegraphics[width=\linewidth,trim={0 2cm 0 2cm},clip]{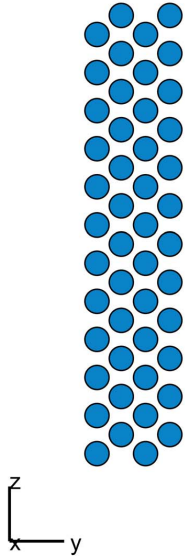}}
\caption{Massive MIMO antenna array geometry. Horizontal element spacing is half-wavelength and vertical element spacing is about 0.7-wavelength.}
\label{fig:array_geometry}
\end{figure}

We consider a macro cellular network environment.
Fig.~\ref{fig:cell_sinr} shows the 3GPP hexagonal cellular network geometry with an inter-site distance of 1.732 km.
With 5-MHz channel bandwidth, 20-W transmission power, 32-m cell tower height, sector antenna pattern with 6\degree\ downtilt,
Fig.~\ref{fig:cell_sinr} plots the typical signal-to-interference-plus-noise ratio (SINR) in a cellular network.
We would like to characterize the wireless channel of a Massive MIMO antenna array in such a macro cellular environment.
Similar to the currently deployed cellular antenna form factor, Fig.~\ref{fig:array_geometry} illustrates a staggered 48-element antenna array with 4 columns where each column has 12 antenna elements.
We focus on the mid-band of 2 GHz typical for macro cell coverage, and the antenna array has half-wavelength element spacing horizontally and approximately 0.7-wavelength element spacing vertically.

\begin{figure}[htbp]
\centerline{\includegraphics[width=\linewidth,trim={0 10cm 0 0cm},clip]{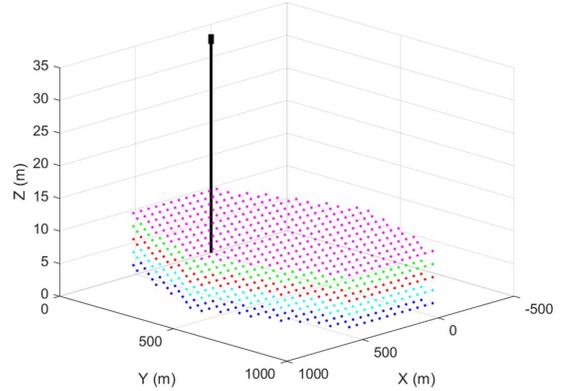}}
\caption{Typical macro cell tower with local scatterers near the ground.}
\label{fig:cell_tower}
\end{figure}

Consider the sector where such a Massive MIMO antenna array is deployed, which is illustrated in Fig.~\ref{fig:cell_tower}.
We assume there are no scatterers high above the ground around the cell tower, but there may be scatterers around the mobile users near the surface of the ground.
In our example, we assume the local scatterers are distributed within the hexagonal sector boundary up to 10 m high.
Suppose we observe the transmit signal at $L$ observation locations in the cell sector.
With interference from the other sectors treated as noise, the wireless channel from the Massive MIMO antenna array to the cell sector is characterized by an $L\times N$ complex baseband matrix $H[k] \in \mathbb{C}^{L\times N}$ for each subband $k$, with $N$ being the number of elements on the antenna array.
We assume there are $K$ subbands over the bandwidth of the wireless channel and thus $1\leq k\leq K$.
We have $N=48$ for the antenna array under consideration in Fig.~\ref{fig:array_geometry}.
In Fig.~\ref{fig:cell_tower}, we consider $L=1730$ regularly spaced observation points in the cell sector (in the three-dimensional space between 0 m and 10 m).

\subsection{Digital vs.\ Hybrid Massive MIMO}

We focus on the downlink of a Massive MIMO system.
In general, a Digital Massive MIMO system may be described by
\begin{IEEEeqnarray}{rCl}
y_D[k] & = & H[k]D[k]x[k]+z[k], \label{eq:digital_massive_mimo}
\end{IEEEeqnarray}
where $k=1,\ldots,K$ is the subband index, $x[k] \in \mathbb{C}^S$ are the $S$ independent complex baseband source symbols,
$D[k] \in \mathbb{C}^{N\times S}$ is the digital precoder,
$H[k] \in \mathbb{C}^{L\times N}$ is the channel,
$y_D[k] \in \mathbb{C}^L$ are the observed signals at the given $L$ locations in the sector,
and $z[k] \in \mathbb{C}^L$ is the additive receiver noise.
The above formulation explicitly considers the observed signals at $L$ locations.
Typically, the system will serve a small number of $\bar{U}$ users out of the $L$ locations.
A characteristic of Massive MIMO is that $N\gg S=\bar{U}$.
For example, current commercial Massive MIMO systems typically target $N=32 \text{ to } 64$ antenna elements per polarization,
with $S=\bar{U}=4 \text{ to } 8$ independent source data streams or users.
Conceptually, a Digital Massive MIMO system may have an RF chain, or digital-to-analog converter (DAC) and analog-to-digital converter (ADC) pair, associated with each antenna element.
Therefore, it has the most flexibility and is able to apply any (linear) precoding $D[k]$ from the inputs to the antenna elements.
Assuming the transmitter has knowledge of the channel matrix $H[k]$ for the subbands $k=1,\ldots,K$, either through reciprocity in time-division duplexing (TDD) or feedback in frequency-division duplexing (FDD), the system may have a subband-specific precoder $D[k]$ for each subband $k$.

On the other hand, a Hybrid Massive MIMO system is described by
\begin{IEEEeqnarray}{rCl}
y_H[k] & = & H[k]WF[k]x[k]+z[k], \label{eq:hybrid_massive_mimo}
\end{IEEEeqnarray}
where $x[k], H[k], y_H [k], z[k]$ have similar interpretation as before,
$F[k] \in \mathbb{C}^{M\times S}$ is the digital precoder,
$W \in \mathbb{C}^{N\times M}$ are the analog beamforming coefficients, and $M$ is the number of RF chains in the system.
In a Hybrid Massive MIMO system, for implementation efficiency, the number of RF chains is much smaller than the number of antenna elements:
$N\gg M=S$.
Moreover, even though the beamforming coefficients $W$ may be digitally controlled, they are typically realized by analog attenuators and phase shifters.
Therefore, $W$ is common across all subbands (thus notated without the subband index $[k]$).
At first glance, it may appear Hybrid Massive MIMO (\ref{eq:hybrid_massive_mimo}) is inherently less flexible than (\ref{eq:digital_massive_mimo}) and not able to perform subband beamforming.
In the next section, we describe how a Hybrid Massive MIMO system may mimic its digital counterpart to perform subband beamforming in a macro cellular environment.

\section{Channel Degrees of Freedom}

\subsection{Long-Term Channel}

Based on the geometry of the macro cellular sector, we decompose the downlink Massive MIMO channel $H[k]$ as follows:
\begin{IEEEeqnarray}{rCl}
H[k] & = & H_L [k] H_R,\quad k=1,\ldots,K, \label{eq:downlink_massive_mimo}
\end{IEEEeqnarray}
where $H_L[k] \in \mathbb{C}^{L\times L}$ represents the channel induced by local scatterers and thus may vary per subband,
and $H_R \in \mathbb{C}^{L\times N}$ is the line-of-sight channel (we assume there are no scatters around the cell tower) from the antenna array to the observation locations in the cell sector.
Therefore, $H_R$ depends only on the long-term geometry of the cell and we assume it is the same across all subbands (thus the subband index $[k]$ is omitted). Assuming the antenna array front-end is \emph{RF coherent}, i.e., each antenna element has the same amplitude and phase characteristics at the carrier frequency, then the entries of $H_R$ are entirely determined by the angles-of-departure of the observation locations and the long-term channel gains due to propagation path loss \cite{zhang:2015, shahsavari:2018}.
We refer to a Hybrid Massive MIMO system with an RF coherent front end as a \emph{Coherent Hybrid Massive MIMO} system \cite{banu:2017}.

\begin{figure}[htbp]
\centerline{\includegraphics[width=\linewidth,trim={0 10cm 0 0cm},clip]{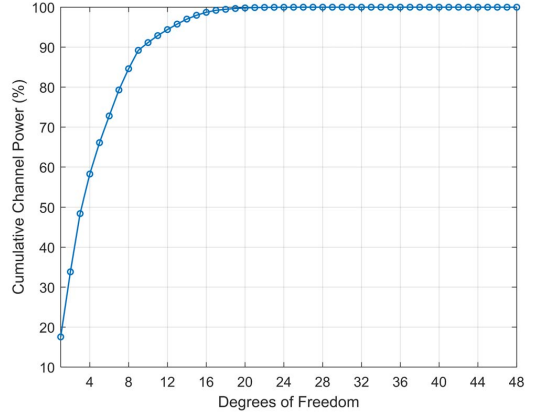}}
\caption{Cumulative power of the descending singular values of the macro cellular channel.}
\label{fig:svd_powers}
\end{figure}

In our problem formulation, $L\gg N$, so the maximum possible rank of $H_R$ is $N$, the number of antenna elements.
However, we show that the effective rank of $H_R$ is much less than $N$ due to the geometry of the macro cellular network.
Based on the dimensions described in Section~\ref{sec:channel_model}, we computed the singular value decomposition (SVD) of $H_R$, and the cumulative power of the channel singular values (in descending order) are plotted in Fig.~\ref{fig:svd_powers}.
In our example, we have $N=48$ antenna elements, but the four largest singular values of $H_R$ capture approximately 60\% of the channel power, while the eight largest ones capture 85\%.
Therefore, a macro cellular channel has limited degrees of freedom.
Based on the typical cellular geometry, 4 to 8 data streams effectively capture most of the available degrees of freedom.
Moreover, once the rank is constricted by the long-term channel $H_R$, it will not be increased subsequently after the signals go through the local-scatterers-induced channel $H_L[k]$.

Motivated by Fig.~\ref{fig:svd_powers}, we consider the low-rank approximation of the long-term channel $H_R\approx \tilde{H}_R(\bar{R})$,
where $\tilde{H}_R(\bar{R})$ is the best approximation of $H_R$ with rank $\bar{R}$.
With SVD $H_R=U\Sigma V^H$, $UU^H=I_L$, $VV^H=I_N$, the low-rank approximation can be obtained by
\begin{IEEEeqnarray}{rCl}
\tilde{H}_R & = & U_t \Sigma_t V_t^H, \label{eq:low_rank_channel_svd}
\end{IEEEeqnarray}
where $U_t \in \mathbb{C}^{L\times t}$, $\Sigma_t \in \mathbb{C}^{t\times t}$, $V_t^H \in \mathbb{C}^{t\times N}$
respectively have the $t$ columns, entries, rows of
$U, \Sigma, V^H$ corresponding to the $t=\bar{R}$ largest singular values of $H_R$.

\subsection{Hybrid Subband Beamforming}

Under the low-rank channel approximation, consider $S=\bar{R}$ in the Digital Massive MIMO System (\ref{eq:digital_massive_mimo}):
\begin{IEEEeqnarray}{rCl}
y_D[k] & =& H_L[k] \tilde{H}_R D[k] x[k]+z[k] \label{eq:y_Dk}\\
 & = & H_L[k] U_t \Sigma_t V_t^H D[k] x[k]+z[k] \label{y_Dk_svd}\\
 & = & H_L[k] U_t \Sigma_t \tilde{D}[k] x[k]+z[k], \label{y_Dk_svd_tilde}
\end{IEEEeqnarray}
where $\tilde{D}[k] \in \mathbb{C}^{S\times S} \triangleq V_t^H D[k]$ is the \emph{effective digital precoder}.
Similarly, under the low-rank channel, consider $M=S=\bar{R}$ in the Hybrid Massive MIMO System (\ref{eq:hybrid_massive_mimo}):
\begin{IEEEeqnarray}{rCl}
y_H[k] & = & H_L[k] \tilde{H}_R W F[k] x[k] + z[k] \label{eq:y_Hk}\\
 & = & H_L[k] U_t \Sigma_t V_t^H W F[k] x[k]+z[k] \label{eq:y_Hk_svd}\\ 
 & = & H_L[k] U_t \Sigma_t F[k] x[k]+z[k] \label{eq:y_Hk_svd_eigen}\\
 & = & y_D[k] \label{eq:y_Hk_y_Dk},
\end{IEEEeqnarray}
where (\ref{eq:y_Hk_svd_eigen}) follows from setting $W=V_t$ to obtain $V_t^H V_t=I_S$,
and (\ref{eq:y_Hk_y_Dk}) follows from setting $F[k]=\tilde{D}[k]$ for all subbands $k=1,\ldots,K$.

Therefore, if the wireless environment has limited degrees of freedom,
which we see is a good approximate model for macro cellular due to its geometry,
the Coherent Hybrid Massive MIMO system is able to emulate the Digital Massive MIMO system subband-by-subband as observed by all locations over the cell.
Furthermore, if the channel degrees of freedom are larger than the number of RF chains $M$ in the Coherent Hybrid Massive MIMO system, the beam coefficients $W$ may be dynamically updated every subframe, effectively allowing the system to time-multiplex digital beamforming $F[k]$ to portions of the cell at a time.

\subsection{Macro Cellular Eigenbeams}

\begin{figure}[htbp]
  \centering
  \subfloat[][Beam pattern]{
     \includegraphics[height=5cm,trim={2cm 11cm 2cm 1cm},clip]{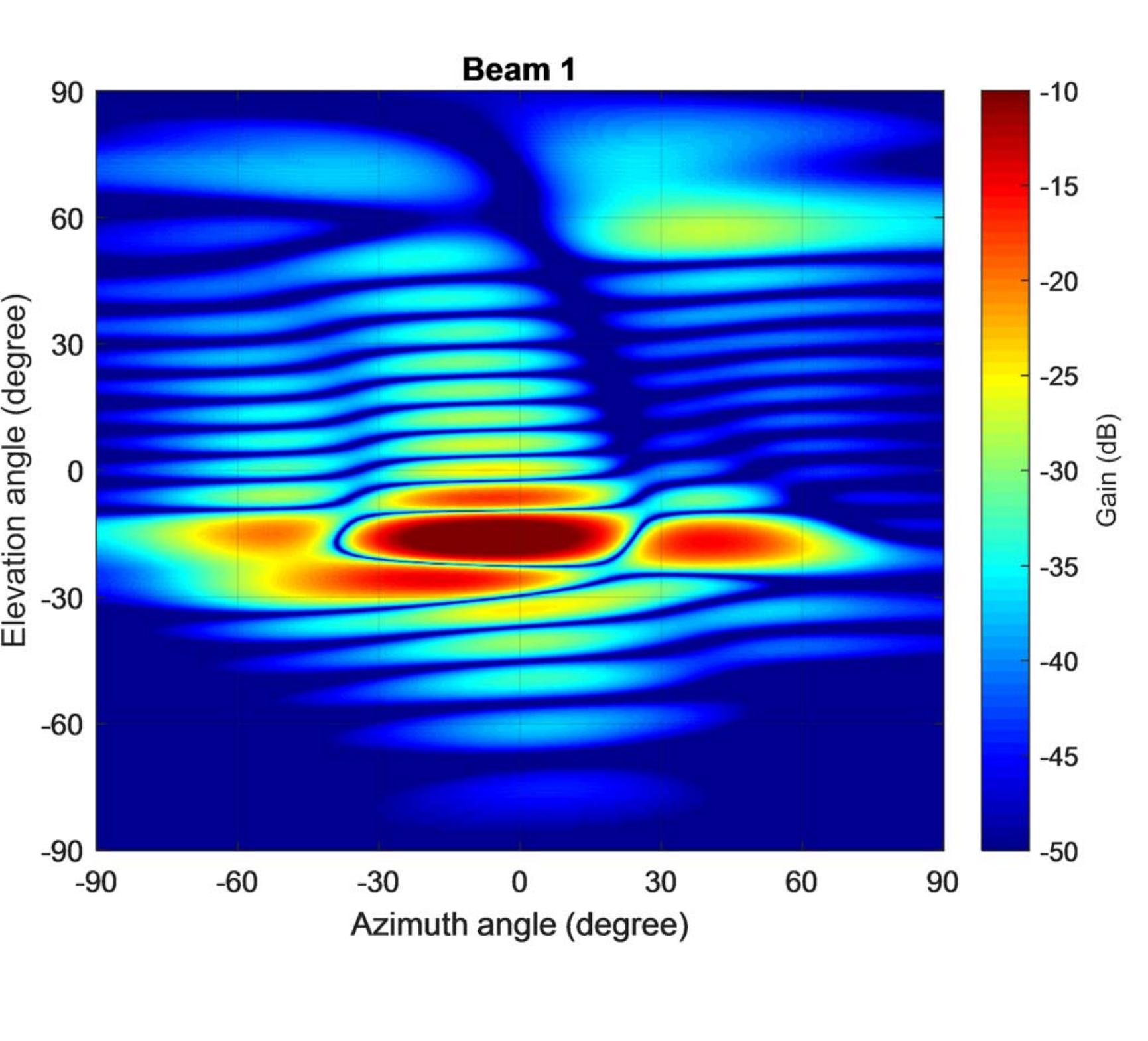}
  }
  \subfloat[][Array power]{
    \includegraphics[height=5cm,trim={7cm 11cm 7cm 1cm},clip]{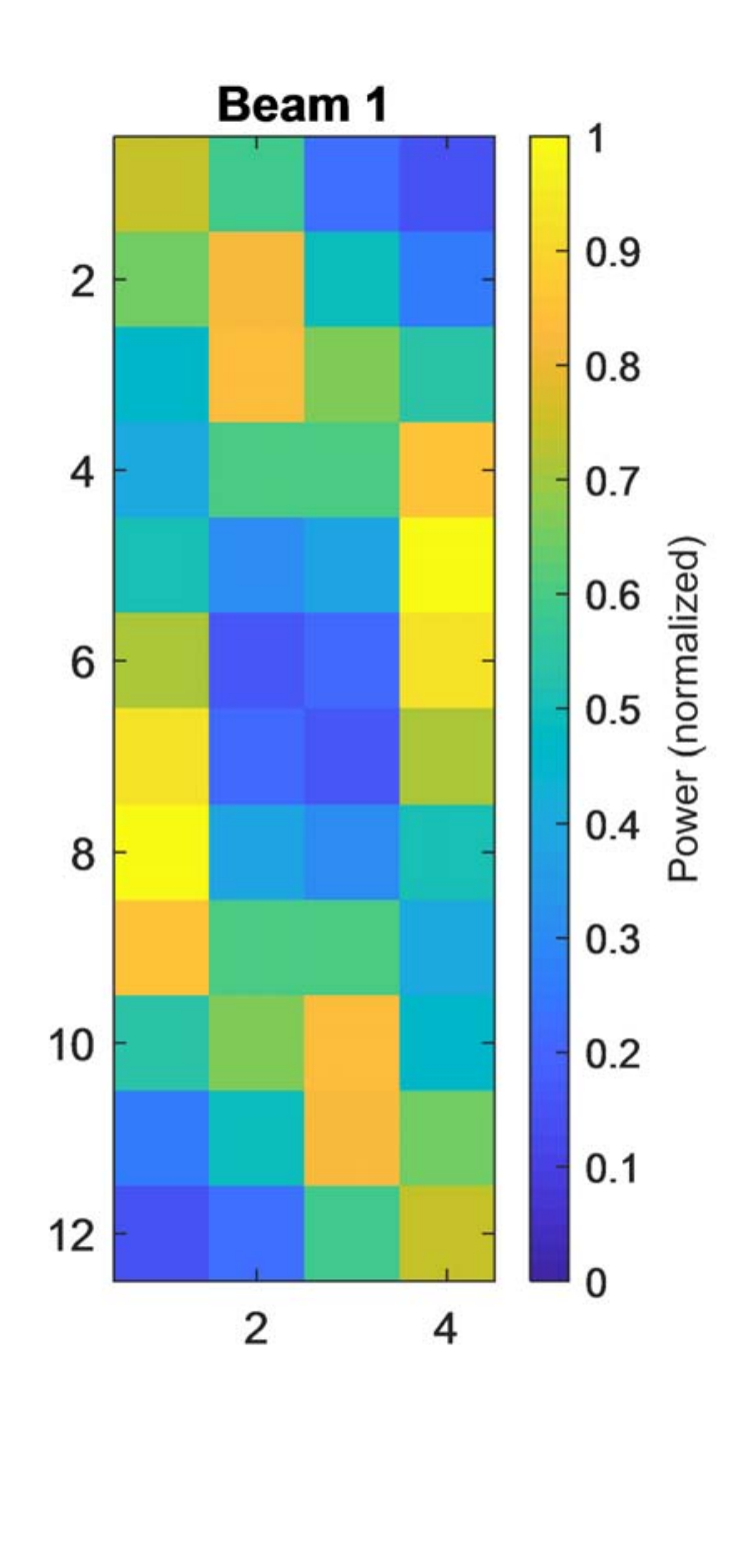}
  }\\
  \subfloat[][Beam pattern]{
     \includegraphics[height=5cm,trim={2cm 11cm 2cm 1cm},clip]{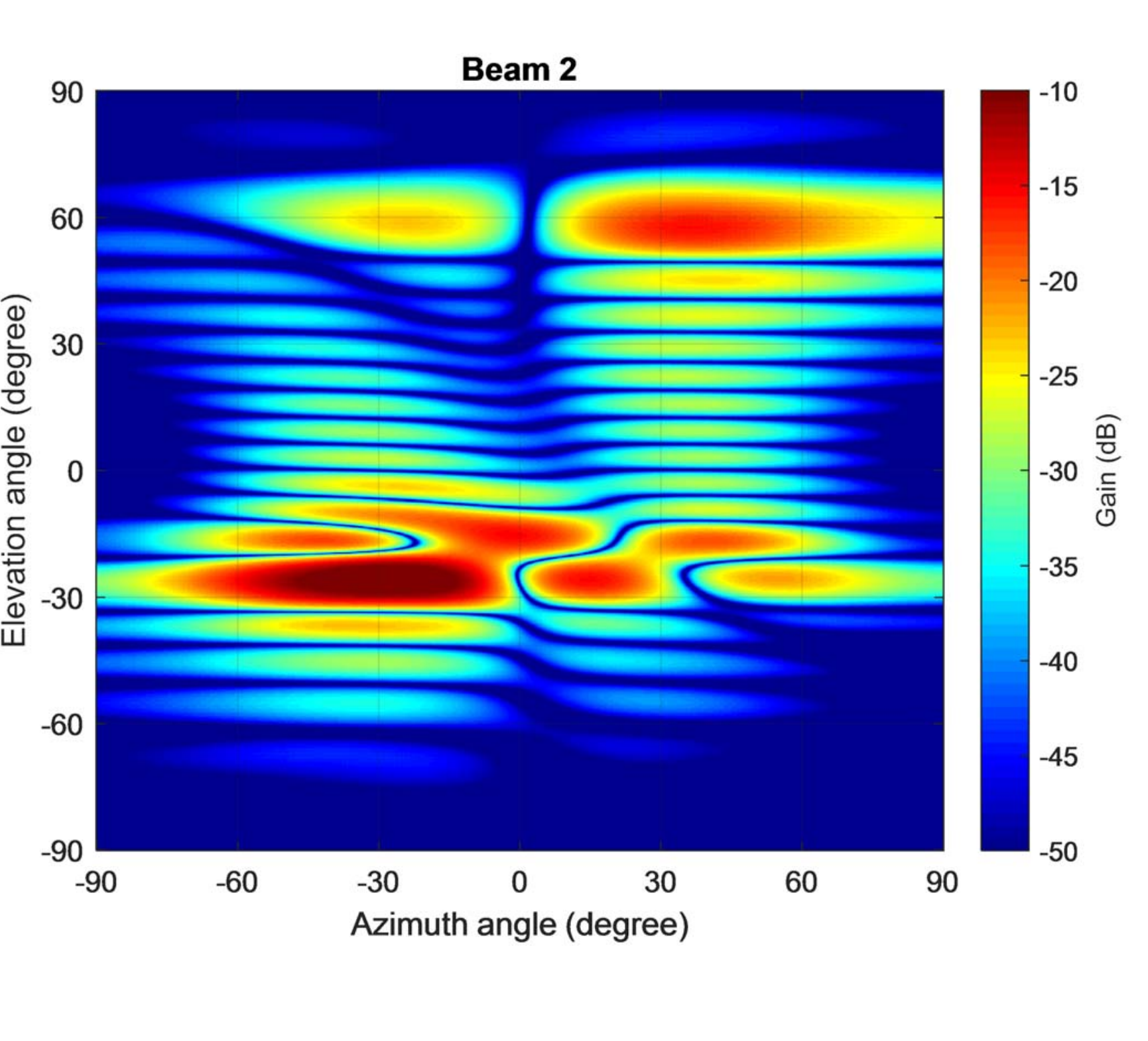}
  }
  \subfloat[][Array power]{
    \includegraphics[height=5cm,trim={7cm 11cm 7cm 1cm},clip]{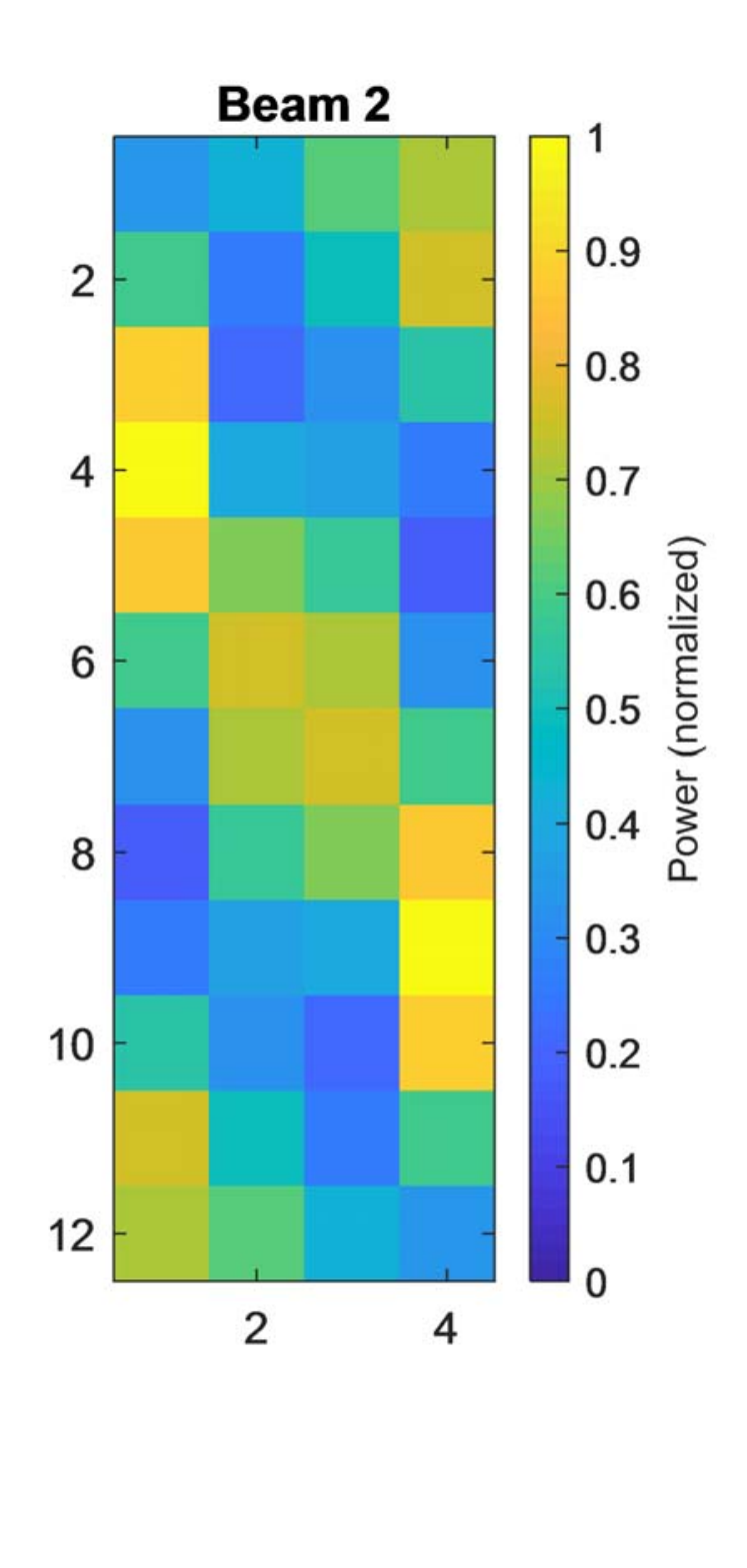}
  }\\
  \subfloat[][Beam pattern]{
     \includegraphics[height=5cm,trim={2cm 11cm 2cm 1cm},clip]{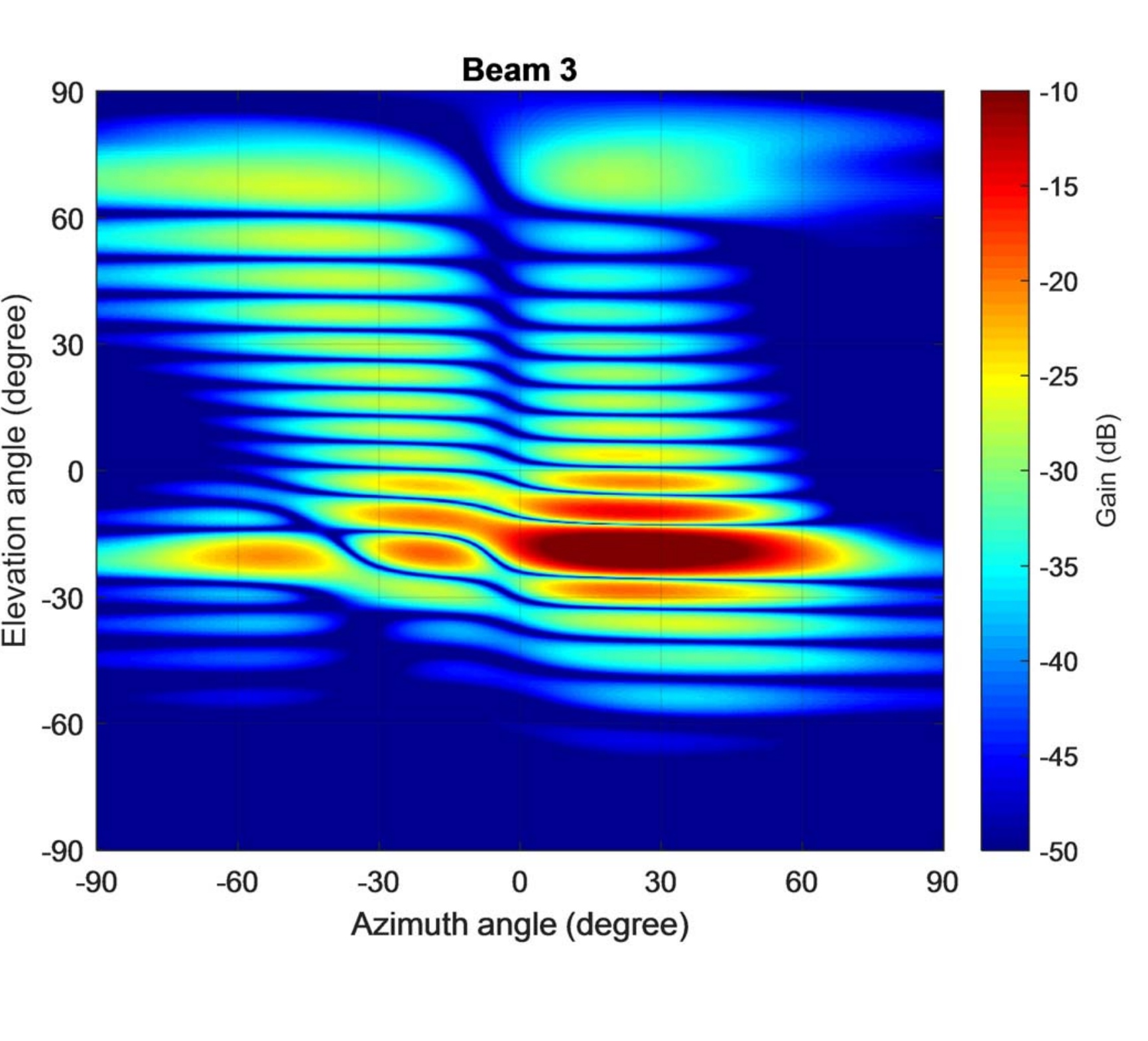}
  }
  \subfloat[][Array power]{
    \includegraphics[height=5cm,trim={7cm 11cm 7cm 1cm},clip]{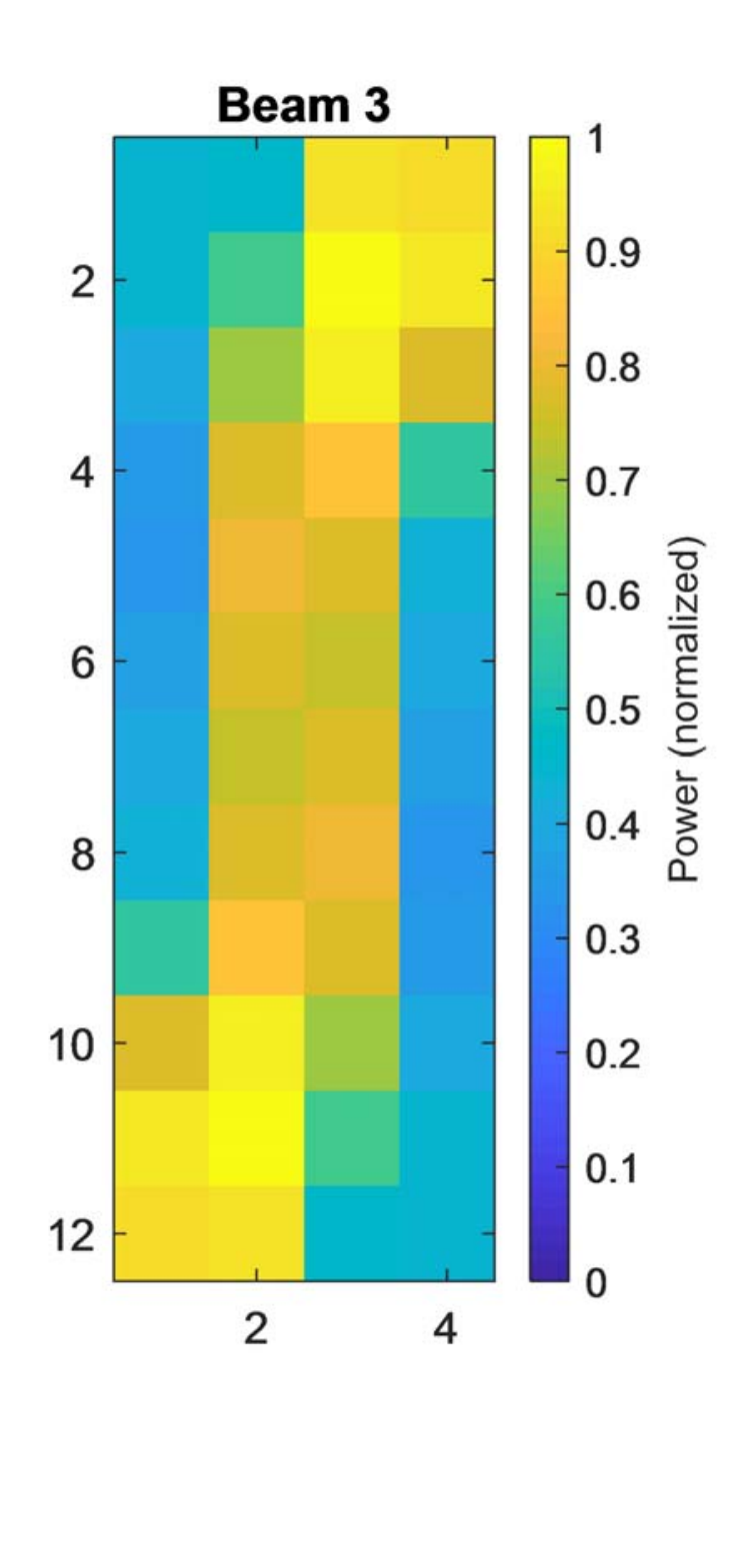}
  }\\
  \subfloat[][Beam pattern]{
     \includegraphics[height=5cm,trim={2cm 11cm 2cm 1cm},clip]{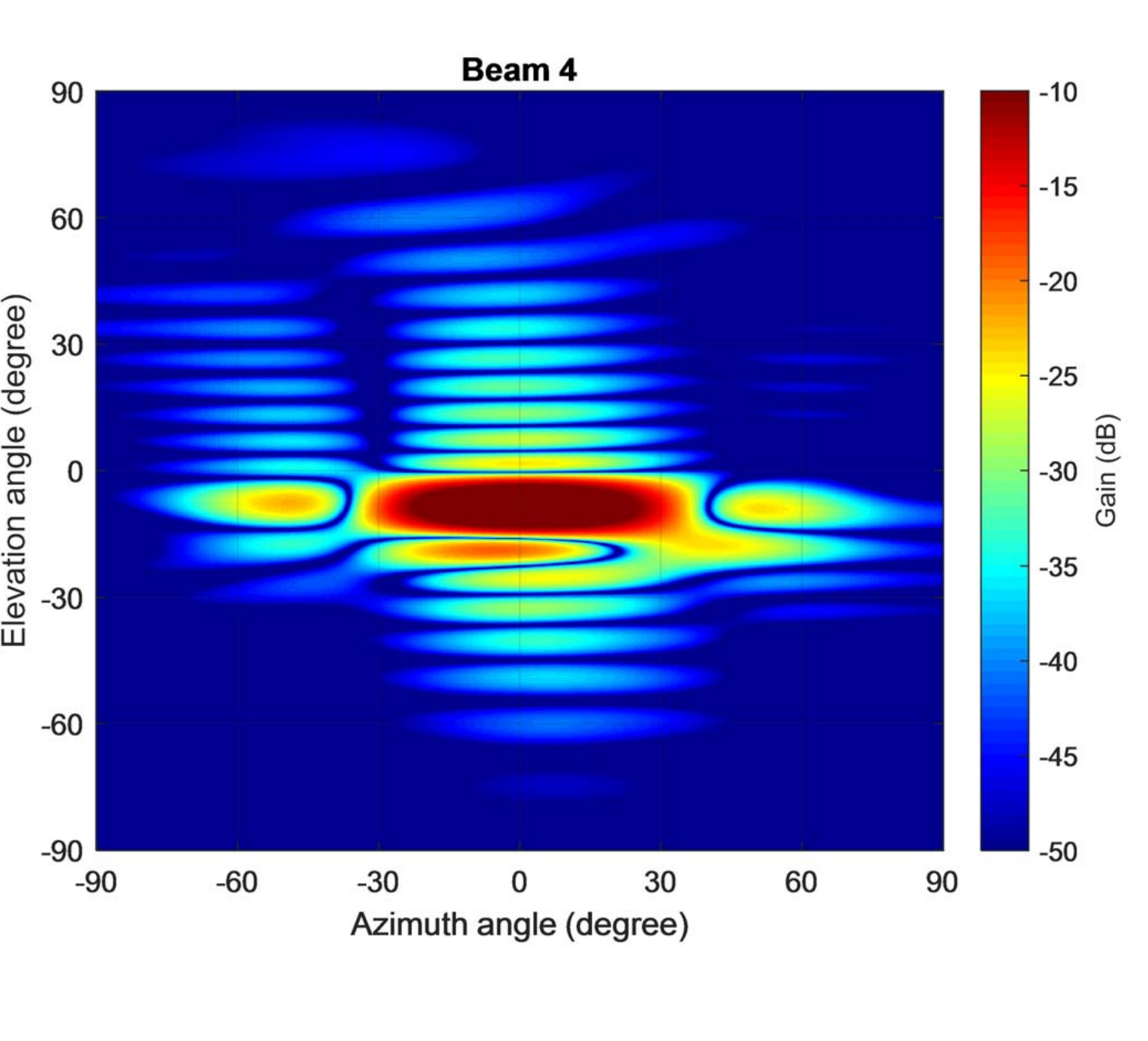}
  }
  \subfloat[][Array power]{
    \includegraphics[height=5cm,trim={7cm 11cm 7cm 1cm},clip]{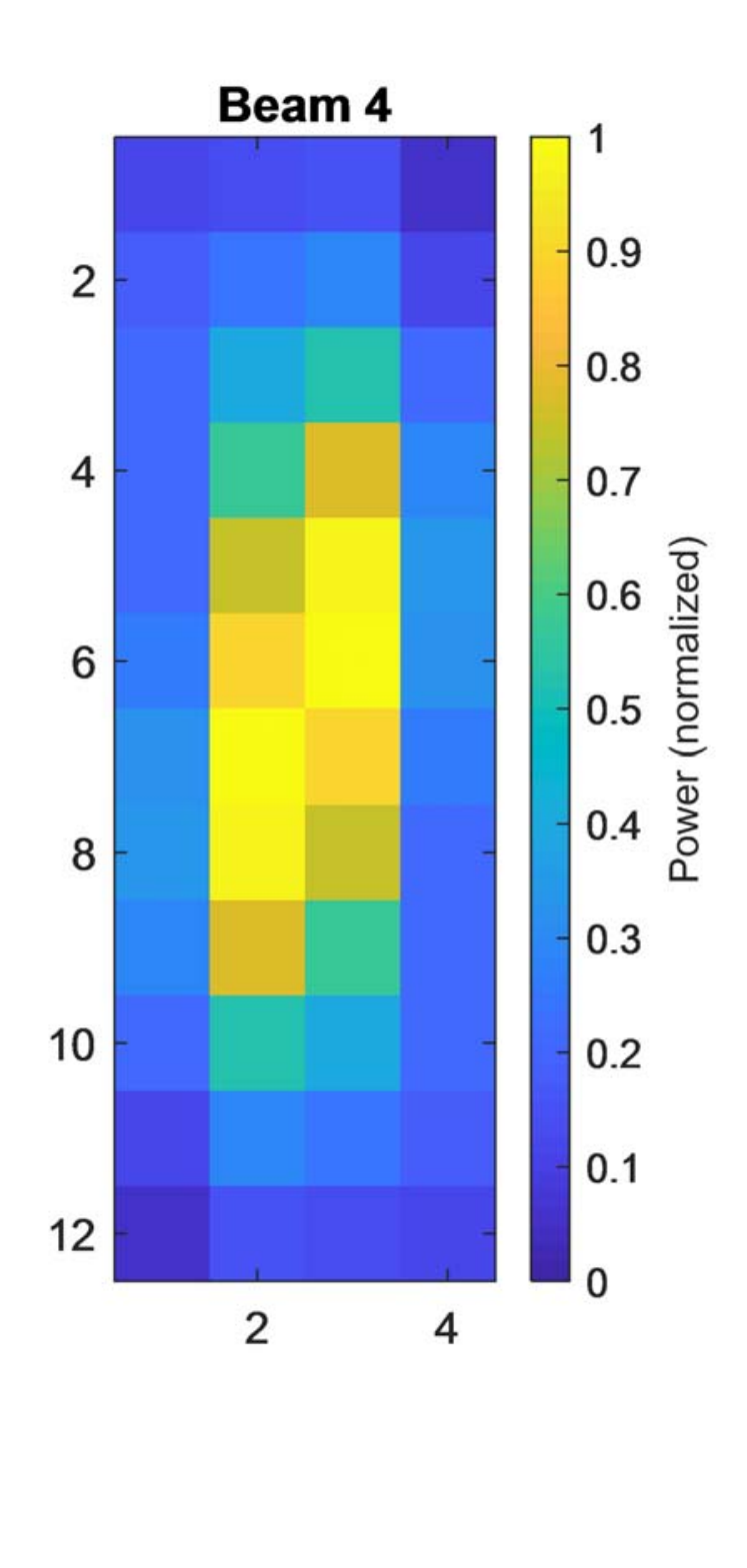}
  }
	\caption{Macro cellular first four eigenbeam patterns and array power distributions.}
  \label{fig:eigenbeams}
\end{figure}

In setting the beam coefficients $W=V_t$ in (\ref{eq:y_Hk_svd}), note that $V_t$ corresponds to the eigenvectors of the covariance matrix
$\tilde{H}_R^H \tilde{H}_R \in \mathbb{S}_{++}$ of the rank-limited long-term channel,
where $\mathbb{S}_{++}$ denotes the set of positive definite Hermitian matrices.
Therefore, we call such set of beam coefficients the \emph{eigenbeams} corresponding to the long-term channel $\tilde{H}_R$.
As a numerical example, for the cellular geometry in Fig.~\ref{fig:cell_tower}, the first four eigenbeams are illustrated in Fig.~\ref{fig:eigenbeams}.
The left sub-figures show the beam patterns with respect to azimuth (horizontal) and elevation (vertical) angles,
and the right sub-figures show the normalized element amplifier output power in the $12 \times 4$ antenna array.
The eigenbeams are a set of beam patterns where a linear combination of them (as determined by the digital precoder $F[k]$ in each subband) mimics the Digital Massive MIMO operation as received at the observation locations.
The eigenbeam patterns can be seen to cover different regions of the sector, while they light up different groupings of amplifiers on the antenna array.
Interestingly, composing linear combinations of eigenbeams to form any arbitrary beam pattern (up to the degrees of freedom of the channel) is analogous to the idea of using ``eigenfaces'' for facial recognition \cite{turk:1991}.

\section{Conclusion}

Hybrid Massive MIMO is viewed as an efficient architecture that helps to address concerns on Massive MIMO implementation complexity.
However, since the beamforming coefficients are realized using analog circuitry, it was perceived that subband beamforming is beyond the capability of Hybrid Massive MIMO systems.
In this paper, we show that a macro cellular channel is limited in its degrees of freedom.
Taking advantage of such property, the long-term component of the channel can be decomposed into a set of eigenbeams, which is independent of the subbands.
A Coherent Hybrid Massive MIMO system can then form arbitrary linear combinations of the eigenbeams to mimic Digital Massive MIMO at every subband as observed across all locations in the cell sector.

\bibliographystyle{IEEEtran}
\bibliography{IEEEabrv,massive_mimo}

\end{document}